\documentclass[pdflatex,sn-mathphys-num,iicol]{sn-jnl}
\usepackage{amsmath}
\usepackage{braket}
\usepackage{graphicx}%
\usepackage{multirow}%
\usepackage{amssymb,amsfonts}%
\usepackage{amsthm}%
\usepackage{mathrsfs}%
\usepackage[title]{appendix}%
\usepackage{xcolor}%
\usepackage{textcomp}%
\usepackage{manyfoot}%
\usepackage{booktabs}%
\usepackage{algorithm}%
\usepackage{algorithmicx}%
\usepackage{algpseudocode}%
\usepackage{listings}%
\usepackage{subcaption}%
\usepackage{braket}
\usepackage{booktabs}  
\usepackage{siunitx}
\usepackage{longtable}
\begin{document}


\title[Direct Calculation of the Two-Neutrino Double-Beta Decay Rate Including the Full Lepton-Energy Dependence]{Direct Calculation of the Two-Neutrino Double-Beta Decay Rate Including the Full Lepton-Energy Dependence}

\author*[1,2,3]{\fnm{S. A.} \sur{Ghinescu}}\email{stefan.ghinescu@nipne.ro}
\author[1,2]{\fnm{A.} \sur{Neacsu}}\email{nandrei@theory.nipne.ro} 
\author*[1]{\fnm{S.} \sur{Stoica}}\email{sabin.stoica@cifra-c2unesco.ro}

\affil[1]{\orgname{International Center for Advanced Training and Research in Physics} \orgaddress{\street{P.O. Box MG-12} \city{M\u{a}gurele}, \postcode{077125},  \country{Romania}}}

\affil[2]{\orgname{Horia Hulubei National Institute of Physics and Nuclear Engineering} \orgaddress{\street{30 Reactorului, POB MG-6} \city{M\u{a}gurele}, \postcode{077125},  \country{Romania}}}

\affil[3]{\orgname{Department of Physics, University of Bucharest} \orgaddress{\street{405 Atomi\c{s}tilor} \city{M\u{a}gurele}, \postcode{077125},  \country{Romania}}}


\abstract{The calculation of the two-neutrino double-beta decay (DBD) rates has so far relied on approximations that decouple the nuclear structure from the emitted lepton kinematics. To ensure a more rigorous treatment, we propose an approach which incorporates the full interdependence between nuclear structure and lepton kinematics within a unified formula and performs calculations without resorting to these approximations. Deviations of the decay rates and electron spectra from the traditional methods, such as closure, non-closure, and Taylor expansion approximations, are presented and discussed for the isotopes $^{82}$Se and $^{136}$Xe. Our approach gives a more realistic description of the DBD process and paves the way for further theoretical and experimental investigations into the correlated nuclear structure interplay with emitted lepton kinematics contributing to the dynamics of the process. Extensions of this framework to other isotopes and to neutrinoless double-beta decay are currently underway.}

\maketitle

\section{Introduction}

Double-beta decay is one of the most promising research topics in physics, as it can provide key information on still unsolved fundamental problems, such as matter-antimatter asymmetry, still unknown properties of neutrinos, and scenarios of physics beyond the Standard Model (BSM). While the primary goal of the DBD experiments remains the detection of the neutrinoless decay mode ($0\nu\beta\beta$), a hypothetical BSM process, the conventional two-neutrino decay mode ($2\nu\beta\beta$) is simultaneously the subject of precise measurements, as a critical benchmark for validating nuclear structure models and theoretical predictions of electron spectra~\cite{Haxton-PPNP1984,Doi-PTPS1985,Tomoda_RPP_1991,Suhonen_PR_1998, Vogel_ARNPS_2002,Avignone_RMP_2008,Vergados_RPP_2012,Dolinski_ARNPS_2019,Blaum-RMP2020,Bossio-JPG2024}.

Traditionally, due to computing limitations, the $2\nu\beta\beta$ decay rate has been evaluated by factorizing it into two independent components, the nuclear matrix elements (NMEs), describing the nuclear structure, and phase-space factors (PSFs) containing the lepton kinematics and atomic effects, respectively. This factorization is achieved by using approximations within the so-called closure~\cite{Haxton-PPNP1984,Doi_PTP_1983,Tomoda_RPP_1991,Suhonen_PR_1998,Vogel_ARNPS_2002}, non-closure~\cite{Horoi_PRC_2007,Novario_PRL_2021,Hinohora_PRC_2022} (N.C.) and Taylor expansion methods~\cite{Simkovic_PRC_2018}. In the closure approximation, in the denominator of the NME expression, the energies of the nuclear $1^+$ states ($E_N$) in the intermediate nucleus are replaced by an average energy $\langle E_N\rangle$, and the lepton energies are approximated by half of the total available energy released. In this way, the summation over the $1^+$ states remains only in the numerator and becomes a sum over a complete set of states, which is done by closure. Since the lepton energies are limited by the Q-value to a few MeV, the decay rate is  sensitive to the exact position of the $1^+$ states, particularly to those with the lowest energies. Consequently, the error in choosing the $\langle E_N\rangle$ value significantly affects the NME calculation and, subsequently, the estimation of the half-life, making the closure approximation inadequate for NME calculations~\cite{PhysRevC.87.045501,engel_status_2017,PhysRevC.87.014315}.  It should be noted that, unlike the $2\nu\beta\beta$ mode,  the dependence of the NME calculation on the $\langle E_N\rangle$ is weak for the $0\nu\beta\beta$ mode, because of presence of the neutrino momentum (on the order of $\sim$100MeV) in the denominator and the contribution of other multipolarities, which justifies the closure approximation for this decay mode. 
Subsequent works use advanced computational techniques to calculate NMEs without approximating the $E_N$ energies, and therefore without relying on the closure (completeness) relation, but still keeping the nuclear part decoupled from the lepton kinematics. However, the decay rate predictions still remain uncertain, and the accuracy of the NME calculations, which is the main source of uncertainty, can be evaluated only by comparison with the experimental half-lives. A more recent work~\cite{Simkovic_PRC_2018} improved the derivation of the decay rate by retaining the lepton energies in the denominator up to the fourth order in a Taylor expansion. This formalism introduces new products of NMEs and their corresponding PSFs in the decay rate formula. This method was used in several subsequent works for new theoretical predictions and experimental analyses regarding the nuclear effects and electron spectra deviations, due to the additional new terms~\cite{Nitescu_Universe_2021, Nitescu-U2024b, Nitescu_JPG_2024, Nitescu-PRC2025, NEMO3_EPJC_2025, Ghinescu_EPJC_2026}. Nonetheless, this Taylor-expansion method is computationally intensive and its mathematical validity is questionable for certain isotopes in certain parts of the integration domain.

In this Letter, we address these challenges by proposing a novel method to calculate the \(2\nu\beta\beta\) decay rate that integrates the NMEs and PSFs into a unified formulation. This approach preserves the full interdependence between the nuclear structure and lepton kinematics governing the double-beta decay process. Besides the decay rate, we also compute its differential distributions by taking into account the full coupling between nuclear structure and lepton kinematics. We apply our method to the $2\nu\beta\beta$ decay of the isotopes $^{82}$Se and $^{136}$Xe, leaving the extension to other isotopes and to the $0\nu\beta\beta$ decay mode for future work.
Recently, ref.~\cite{arxiv242114160} has proposed a formulation for the decay rate that resembles ours, but to our knowledge, this is the first time that the effects of nuclear structure on the emitted electrons spectra are investigated.  

\section{Formalism}
Following the formalism presented in~\cite{Simkovic_PRC_2018} and~\cite{Ghinescu_EPJC_2026}, the differential decay rate with respect to the angle between the emitted electrons can be written as
\begin{equation}
    \frac{d\Gamma}{d\cos\theta} = \frac{\Gamma}{2}(1+K\cos\theta)
\end{equation}
where $K$ is the angular correlation coefficient and it is given by~~\cite{Doi_PTP_1983, Haxton-PPNP1984,Doi-PTPS1985, Tomoda_RPP_1991, Doi_PTP_1993}
\begin{equation}
    K = \frac{\Lambda}{\Gamma}
\end{equation}

with $\Gamma$ being total decay rate for the $2\nu\beta\beta$ process and $\Lambda$ is the angular part which can be written as
\begin{align}
\label{eq:gamma}
    \begin{aligned}
        \Gamma &= \frac{\ln 2}{T_{1/2}}\\
               &= g_{A}^{4}\frac{(G_{\beta}|V_{ud}|)^{4}}{8\pi^{7}}\\
        &\times \int_{0}^{Q_{\beta\beta}}d\epsilon_{1}\int_{0}^{Q_{\beta\beta}-\epsilon_{1}}d\epsilon_2\int_{0}^{Q_{\beta\beta}-\epsilon_1-\epsilon_2}d\omega_{1}\\
        &\times p_{1}(\epsilon_{1}+m_e)p_2(\epsilon_{2}+m_e)\omega_{1}^{2}\omega_{2}^{2}\mathcal{A}^{2\nu}f_{11}^{(0)}\\
        \Lambda &= g_{A}^{4}\frac{(G_{\beta}|V_{ud}|)^{4}}{8\pi^{7}}\\
        &\times \int_{0}^{Q_{\beta\beta}}d\epsilon_{1}\int_{0}^{Q_{\beta\beta}-\epsilon_{1}}d\epsilon_2\int_{0}^{Q_{\beta\beta}-\epsilon_1-\epsilon_2}d\omega_{1}\\
        &\times p_{1}(\epsilon_{1}+m_e)p_2(\epsilon_{2}+m_e)\omega_{1}^{2}\omega_{2}^{2}\mathcal{B}^{2\nu}f_{11}^{(1)}
    \end{aligned}
\end{align}
where $T_{1/2}$ is the half-life of the nucleus under this process, $g_{A}$ is the weak axial-vector constant, $G_{\beta}$ is the Fermi constant, $V_{ud}$ is the first element of the Cabibbo-Kobayashi-Maskawa (CKM) matrix, $p_{1,2}=\sqrt{\epsilon_{1,2}(\epsilon_{1,2}+2m_{e})}$ are the momenta of the emitted electrons, $\epsilon_{1,2}$ are the kinetic energies of the emitted electrons, $\omega_{1,2}$ are the energies of the antineutrinos, and $Q_{\beta\beta}$ is the Q-value of the process (the total kinetic energy of the emitted leptons).

The factors $f_{11}^{(0)}$ and $f_{11}^{(1)}$ are given by combinations of the emitted electrons' wave functions; their formula can be found in many works, for example~\cite{Doi_PTP_1983,Haxton-PPNP1984,Doi-PTPS1985,Tomoda_RPP_1991,Kotila_PRC_2012,Stoica-PRC2013}:
\begin{align}
    \begin{aligned}
    \label{eq:f110}
        f_{11}^{0} &= \left|f^{-1-1}\right|^{2}+\left|f_{11}\right|^{2}+ \left|{f^{-1}}_{1}\right|^{2} + \left|{f_1}^{-1}\right|^{2},\\
        f_{11}^{1} &= -2\Re\left(f^{-1-1}f^{*}_{11} + {f^{-1}}_{1}{f_1}^{-1*}\right)
    \end{aligned}
\end{align}
where
\begin{align}
    \begin{aligned}
        f^{-1-1} &= g_{-1}(\epsilon_{1}) g_{-1}(\epsilon_{2}) \hspace {1cm} f_{1,1} = f_{1}(\epsilon_{1})f_{1}(\epsilon_{2})\\
        {f^{-1}}_{1} &= g_{-1}(\epsilon_{1})f_{1}(\epsilon_{2}) \hspace {1cm} 
        {f_{1}}^{-1} = f_{1}(\epsilon_{1})g_{-1}(\epsilon_{2}),\\
    \end{aligned}
\end{align}
where $g_{-1}(\epsilon)\equiv g_{-1}(\epsilon, R)$ and $f_{1}(\epsilon) \equiv f_{1}(\epsilon, R)$ are the large and small components of an s-wave emitted electron in the field of the daughter atom. We compute the $g_{-1}$ and $f_{1}$ functions as explained in~\cite{Nitescu-PRC2023} using the \verb|RADIAL| package~\cite{Salvat-CPC2019} with the modified self-consistent Dirac-Hartree-Fock-Slater treatment of the final atom, thus accounting for screening effects and final nuclear size.

The expressions $\mathcal{A}^{2\nu}$ and $\mathcal{B}^{2\nu}$ include the nuclear structure contribution coupled to the lepton kinematics~\cite{Simkovic_PRC_2018}: 
\begin{align}
\label{eq:a2nu}
    \begin{aligned}
        \mathcal{A}^{2\nu} &= \frac{1}{4}\left(\left|M_{GT}^{K}+M_{GT}^{L}\right|^{2} + \frac{1}{3}\left|M_{GT}^{K}-M_{GT}^{L}\right|^{2}\right)\\
        \mathcal{B}^{2\nu} &= \frac{1}{4}\left(\left|M_{GT}^{K}+M_{GT}^{L}\right|^{2} - \frac{1}{9}\left|M_{GT}^{K}-M_{GT}^{L}\right|^{2}\right)\\
    \end{aligned}
\end{align}
where 
\begin{equation}
\label{eq:MGT}
    M_{GT}^{K,L}=\sum_{N}M_{N}\frac{E_{N}-\frac{1}{2}(E_{I}+E_{F})}{\left[E_{N}-\frac{1}{2}(E_{I}+E_{F})\right]^{2}-\epsilon_{K,L}^{2}},
\end{equation}
with
\begin{equation}
\label{eq:MN}
    M_{N} = \bra{0^{+}_{F}}|\sum_{n}\tau_{n}^{-}\sigma_{n}|\ket{1^{+}_{N}}\bra{1^{+}_{N}}|\sum_{n}\tau_{n}^{-}\sigma_{n}|\ket{0^{+}_{I}}.
\end{equation}
Here $\ket{0_{I,F}^{+}}$ are the ground states of the initial and final nuclei, respectively. The energies of the intermediate $\ket{1_{N}^{+}}$ states are denoted by $E_{N}$, while $E_I$ and $E_F$ denote the energies of the initial and final nuclear states. The operator $\tau^{-}\sigma$ is the Gamow-Teller operator that transforms a neutron into a proton, and the sum in Eq.~\ref{eq:MGT} goes over all possible intermediate $1^{+}$ states. The terms $\epsilon_{K,L}$ are given by
\begin{align}
    \begin{aligned}
        \epsilon_{K}&=\frac{1}{2}\left(\epsilon_{1}+\omega_{1}-\epsilon_{2}-\omega_{2}\right)\\
        \epsilon_{L}&=\frac{1}{2}\left(\epsilon_{1}+\omega_{2}-\epsilon_{2}-\omega_{1}\right)
    \end{aligned}
\end{align}

As already mentioned, historically, the DBD decay rate could not be calculated in this form due to computational limitations related to both the complex summation over the $1^+$ states in the intermediate nucleus and the full interdependence between the nuclear structure and lepton degrees of freedom. To circumvent these challenges, the closure and N.C. approximations were used to factorize it as follows:

\begin{equation}
    \begin{Bmatrix}
            \Gamma\\
            \Lambda
        \end{Bmatrix} = g_{A}^{4} \left|M_{2\nu}\right|^{2}\begin{Bmatrix}
            G^{2\nu}\\
            H^{2\nu}
        \end{Bmatrix}
\end{equation}
and calculate $M_{2\nu}$ (NMEs) and $G_{2\nu}$ or $H^{2\nu}$ (PSFs) as independent factors.

In the Taylor method, an expansion is done around $\epsilon_{K,L}^{2}/\left[E_{N}-\frac{1}{2}(E_{I}+E_{F})\right]^{2} = 0$, and the terms up to the 4th order are retained~\cite{Simkovic_PRC_2018}. The procedure is based on the assumption that the lepton terms, $\epsilon_{K,L}$, cannot exceed $Q_{\beta\beta}/2$ in magnitude, while for most nuclei the terms $E_{N}-\frac{1}{2}(E_{I}+E_{F})$ are larger than $Q_{\beta\beta}/2$ for all excited $1^{+}$ states. In this approach, the decay rate is expressed as a sum of four terms, each term being a product of specific NMEs and its corresponding phase-space factor:

\begin{equation}
    \begin{Bmatrix}
            \Gamma\\
            \Lambda
        \end{Bmatrix} \simeq 
        \begin{Bmatrix}
            \Gamma_{0} + \Gamma_{2}+\Gamma_{22}+\Gamma_{4}\\
            \Lambda_{0} + \Lambda_{2}+\Lambda_{22}+\Lambda_{4}
        \end{Bmatrix}
\end{equation}
where 
\begin{equation}
    \begin{Bmatrix}
            \Gamma_N\\
            \Lambda_N
        \end{Bmatrix}= g_A^{4}|\mathcal{M}_{N}|^{2}\begin{Bmatrix}
            G_N\\
            H_N
        \end{Bmatrix} \hspace{1cm} N=0,2,22,4.
\end{equation}
In the expression above, $\mathcal{M}_{N}$ depends only on the nuclear structure, while $G_{N}$ are phase space factors encompassing only the kinematic effects. The detailed expressions for $\mathcal{M}_{N}$ and $G_{N},H_N$ can be found, for example, in~\cite{Simkovic_PRC_2018} and~\cite{Nitescu_Universe_2021}. Therefore, this method still preserves the decoupling between nuclear and lepton degrees of freedom, and the terms of specific products of NMEs and PSFs are computed separately. This method improved the description of the $2\nu\beta\beta$ decay rate and led to new experimental investigations into nuclear contributions, electron spectra and the determination of the axial-vector constant ($g_A$)~\cite{NEMO3_EPJC_2025,KamLANDZen_PRL_2019,CUPIDMo_PRL_2023,CUORE_PRL_2025,PANDAX_arXiv_2025}, along with theoretical developments in calculating the new $\mathcal{M}_N$, $G_N$ and $H_N$  quantities~\cite{Nitescu_Universe_2021,Nitescu-U2024b,Nitescu_JPG_2024,Nitescu-PRC2025}. However, we note that this method is potentially inapplicable in certain regions of the integration domain. For example, there are isotopes like $^{100}$Mo and $^{82}$Se where the first $1^{+}$ state is just the g.s. of the intermediate nucleus ($^{100}$Tc) or is assigned very close to the g.s. ($^{82}$Br)~\cite{noauthor_livechart_2023}. In these cases, the term $E_{N}-\frac{1}{2}(E_I+E_F)$ is close in magnitude to $Q_{\beta\beta}/2$, and could also be close to $\epsilon_{K,L}$ over certain regions of the integration domain. This also (energetically) implies that the transition through this first $1^+$ excited state might contribute significantly to the decay process.

The method we propose is conceptually simple. Given the set of $M_{N}$ in Eqs.~\ref{eq:MN}, we compute one integral per intermediate state and sum the results to obtain the decay rate, without using any approximation. Although this procedure increases the computational complexity, it remains easily manageable with modern computers, summing over 30-50 excited $1^+$ states. Moreover, preserving the interdependence between nuclear structure and the kinematics of the emitted leptons in the calculation procedure conceptually represents a more realistic description of the DBD process. Next, we apply the method to calculate the electron spectra and angular correlations. The single-electron spectrum is defined as the differential of the decay width with respect to the energy of one electron
\begin{align}
\begin{aligned}
\label{eq:single_spectrum}
    \frac{d\Gamma}{d\epsilon_{1}} &= \frac{g_{A}^{4}(G_{\beta}|V_{ud}|)^{4}}{8\pi^{7}}p_{1}(\epsilon_{1}+m_e)\\
    &\times\int_{0}^{Q_{\beta\beta}-\epsilon_{1}}d\epsilon_2\int_{0}^{Q_{\beta\beta}-\epsilon_1-\epsilon_2}d\omega_{1} p_2(\epsilon_{2}+m_e)\\
    &\times f_{11}^{(0)}(\epsilon_{1},\epsilon_{2})\omega_{1}^{2}\omega_{2}^{2}\mathcal{A}^{2\nu},
\end{aligned}
\end{align}
while the summed electron energy spectrum (the spectrum of the sum of kinetic energies of the emitted electrons) is defined as
\begin{align}
\begin{aligned}
\label{eq:summed_spectrum}
        \frac{d\Gamma}{dT} &= \frac{g_{A}^{4}\left(G_{\beta}|V_{ud}|\right)^{4}}{8\pi^{7}} \frac{T}{Q_{\beta\beta}}\\
        &\times \int_{0}^{Q_{\beta\beta}}dV\int_{0}^{Q_{\beta\beta}-T}d\omega_{1}  \epsilon_1 p_1\epsilon_2 p_2 \\
        &\times f_{11}^{(0)}\omega_{1}^{2}\omega_2^{2}\mathcal{A}^{2\nu},
\end{aligned}
\end{align}
where $T=\epsilon_1 + \epsilon_2$ and $V=Q\epsilon_2/T$. In order to define the angular correlation spectrum, we first define the differential of $\Lambda$ as
\begin{align}
\begin{aligned}
    \label{eq:lambda_spectrum}
     \frac{d\Lambda}{d\epsilon_{1}} &= \frac{g_{A}^{4}(G_{\beta}|V_{ud}|)^{4}}{8\pi^{7}}p_{1}(\epsilon_{1}+m_e)\\
    &\times\int_{0}^{Q_{\beta\beta}-\epsilon_{1}}d\epsilon_2\int_{0}^{Q_{\beta\beta}-\epsilon_1-\epsilon_2}d\omega_{1} p_2(\epsilon_{2}+m_e)\\
    &\times f_{11}^{(1)}(\epsilon_{1},\epsilon_{2})\omega_{1}^{2}\omega_{2}^{2}\mathcal{B}^{2\nu}.
\end{aligned}
\end{align}
The angular correlation spectrum is then given by
\begin{equation}
    \label{eq:alpha_spectrum}
    \alpha(\epsilon_1) = \frac{d\Lambda/d\epsilon_1}{d\Gamma/d\epsilon_1}
\end{equation}

As can be seen from equations~(\ref{eq:single_spectrum},\ref{eq:summed_spectrum},\ref{eq:lambda_spectrum},\ref{eq:alpha_spectrum}) together with equation ~(\ref{eq:a2nu}), the single and summed electron energy spectra as well as the angular correlation spectrum are also affected by the nuclear structure. 

\section{Results}

\begin{table*}[h!]
    \centering
    \begin{tabular}{|c|c|c|c|c|c|c|}
    \hline
        Nucleus & $T_{1/2}^{\text{exp}}$ (yr) & $T_{1/2}^{\text{N.C.}}$ (yr) & $T_{1/2}^{\text{Taylor}}$ (yr) & $T_{1/2}^{\text{Direct}}$ (yr) &  $T_{1/2}^{\text{N.C.}}/T_{1/2}^{\text{Direct}}$ & $T_{1/2}^{\text{Taylor}}/T_{1/2}^{\text{Direct}}$\\
        \hline
       $^{82}$Se & $9.6\times10^{19}$ & $1.19\times10^{19}$ & $1.04\times10^{19}$ & $1.02\times10^{19}$  & 1.163 & 1.017\\
        $^{136}$Xe & $2.2\times10^{21}$ & $4.83\times10^{20}$ & $4.68\times 10^{20}$&$4.67\times10^{20}$ &  1.035 & 1.002\\
       \hline
    \end{tabular}
    \caption{Experimental and theoretical half-lives computed via the Direct method, by the N.C. approximation and by the Taylor expansion method. We have assumed $g_{A}=1.275$ and we have not used any quenching.}
    \label{tab:half_lives_comp}
\end{table*}

\begin{figure*}[h!]
    \centering
    \includegraphics[width=0.45\linewidth]{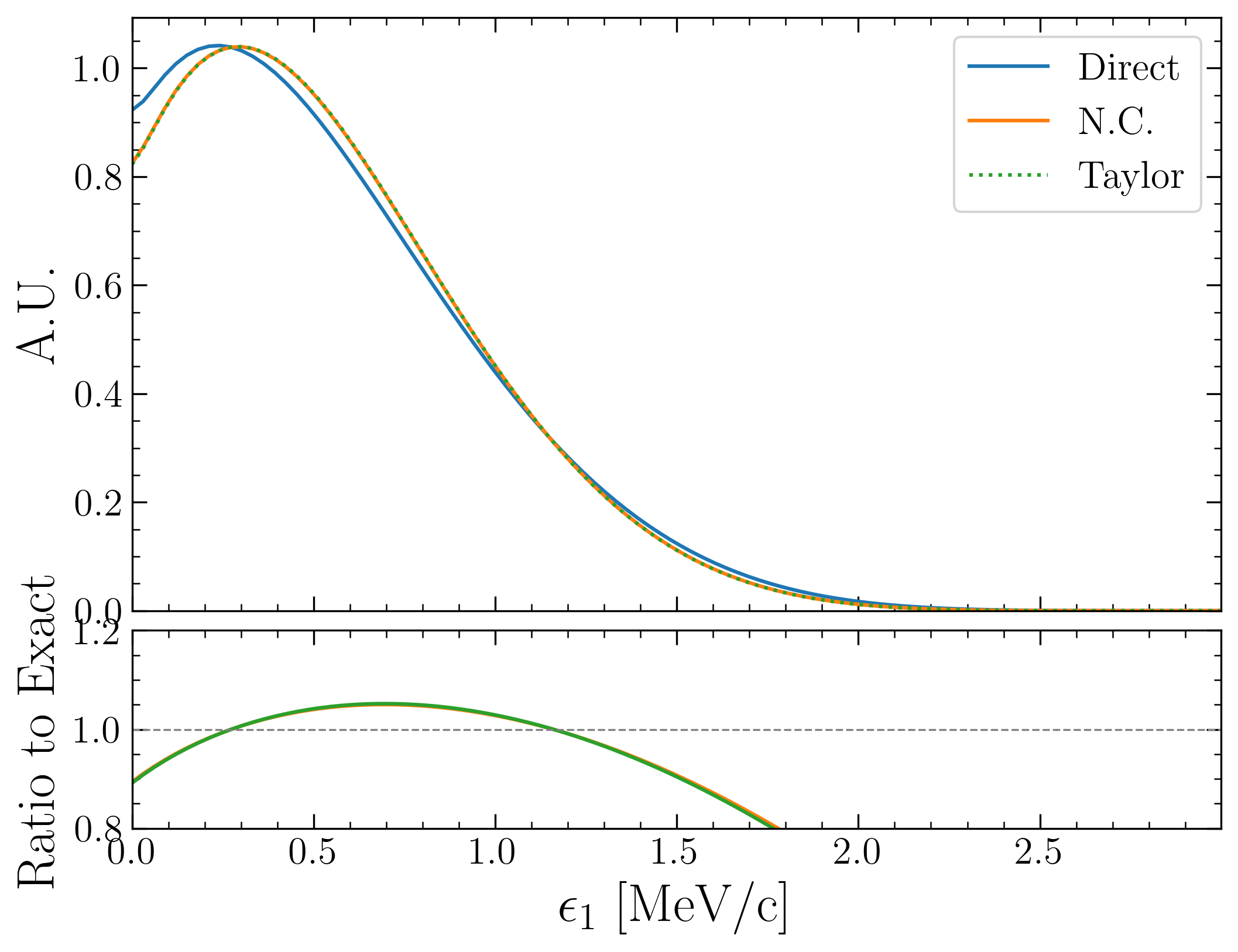}~
    \includegraphics[width=0.45\linewidth]{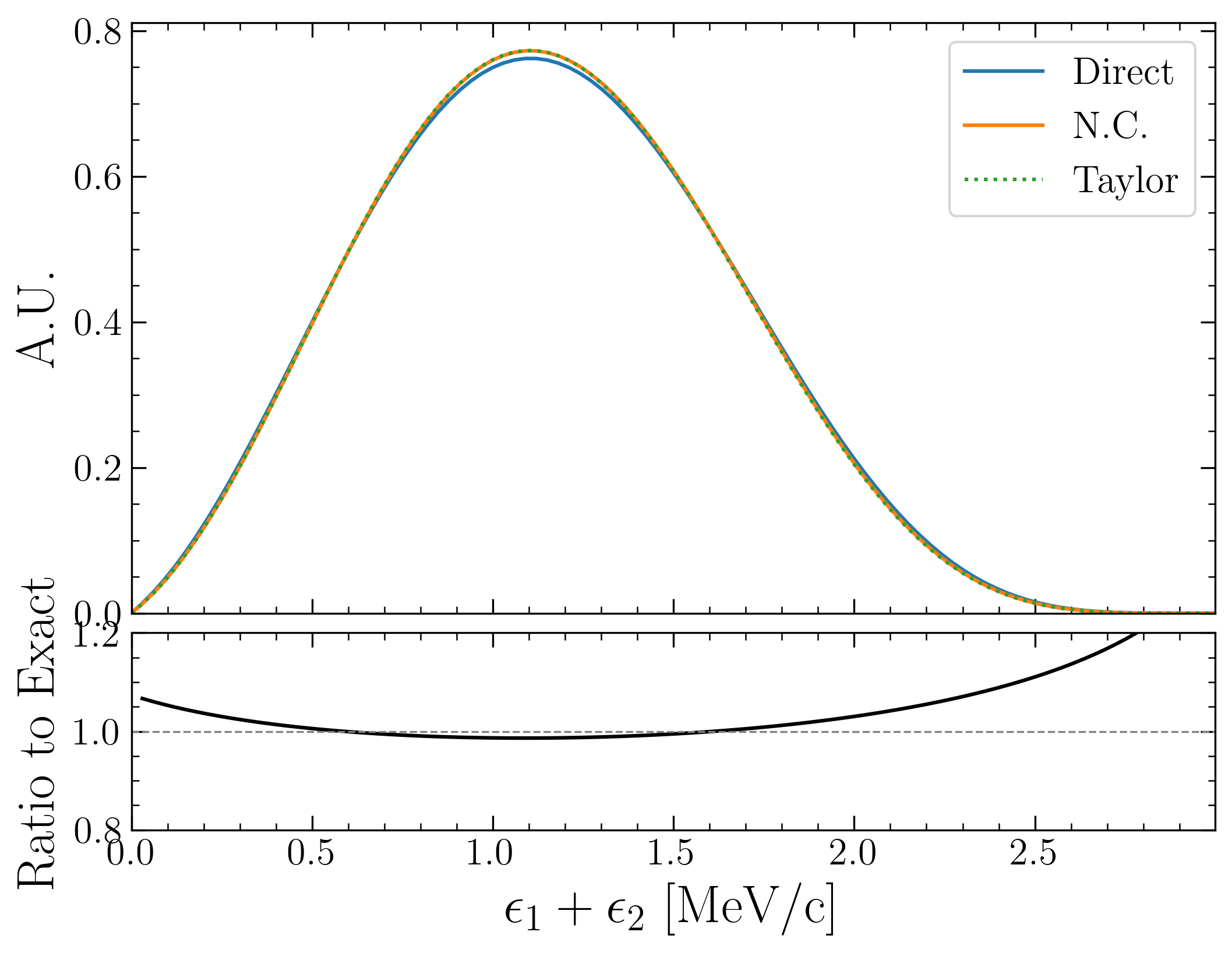}
    \caption{Top: Single (left) and summed (right) electron energy spectra for the $2\nu\beta\beta$ decay of $^{82}$Se. In the upper panels the blue line represents the spectrum obtained with the Direct method, the continuous orange line is the spectrum obtained with the N.C. approximation and the green, the dotted line is the spectrum obtained with the first order Taylor expansion spectrum. In the lower parts we plotted the ratio between approximate spectra and the Direct one.}
    \label{fig:82Se_full_comp}
\end{figure*}

\begin{figure*}[h!]
    \centering
    \includegraphics[width=0.45\linewidth]{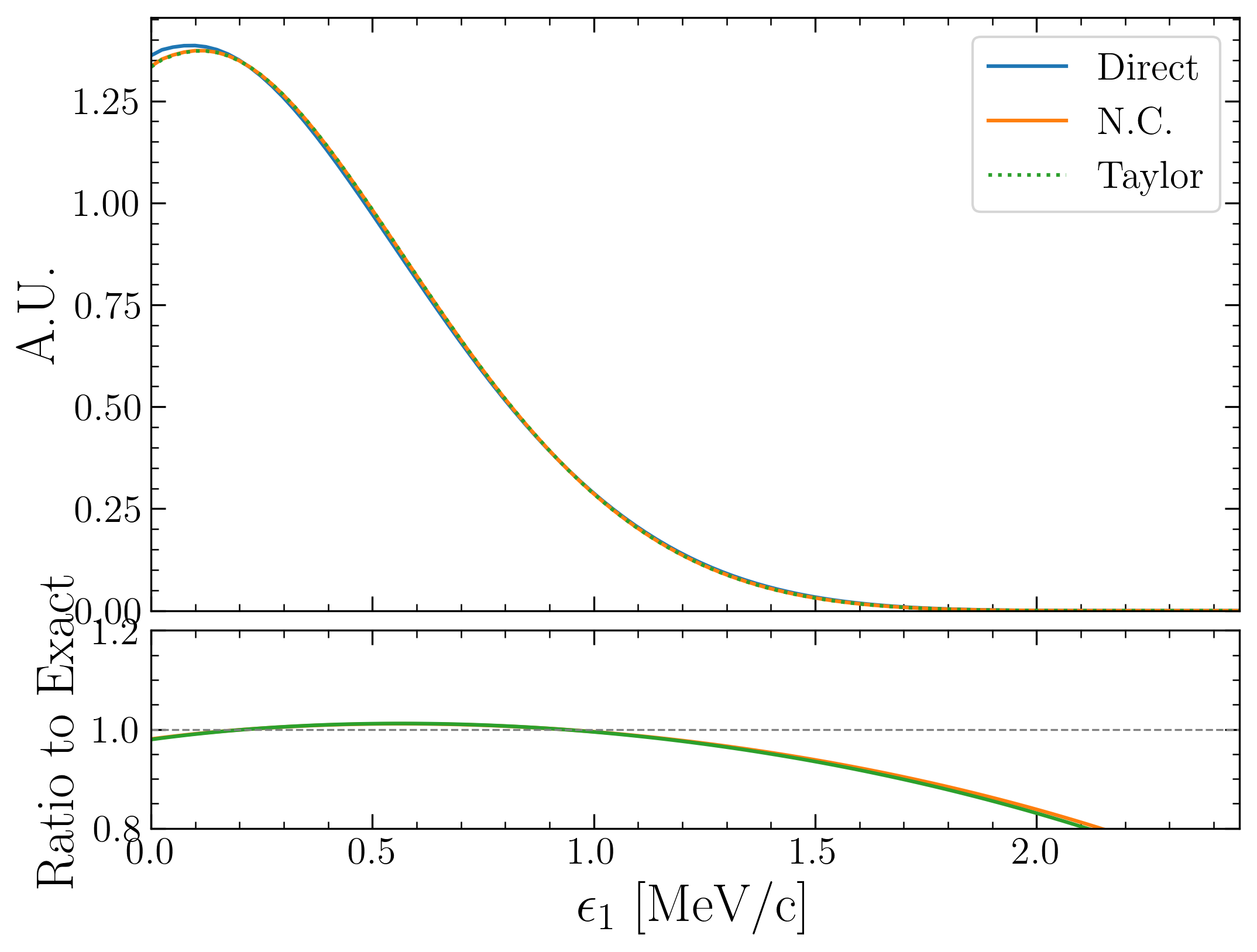}~
    \includegraphics[width=0.45\linewidth]{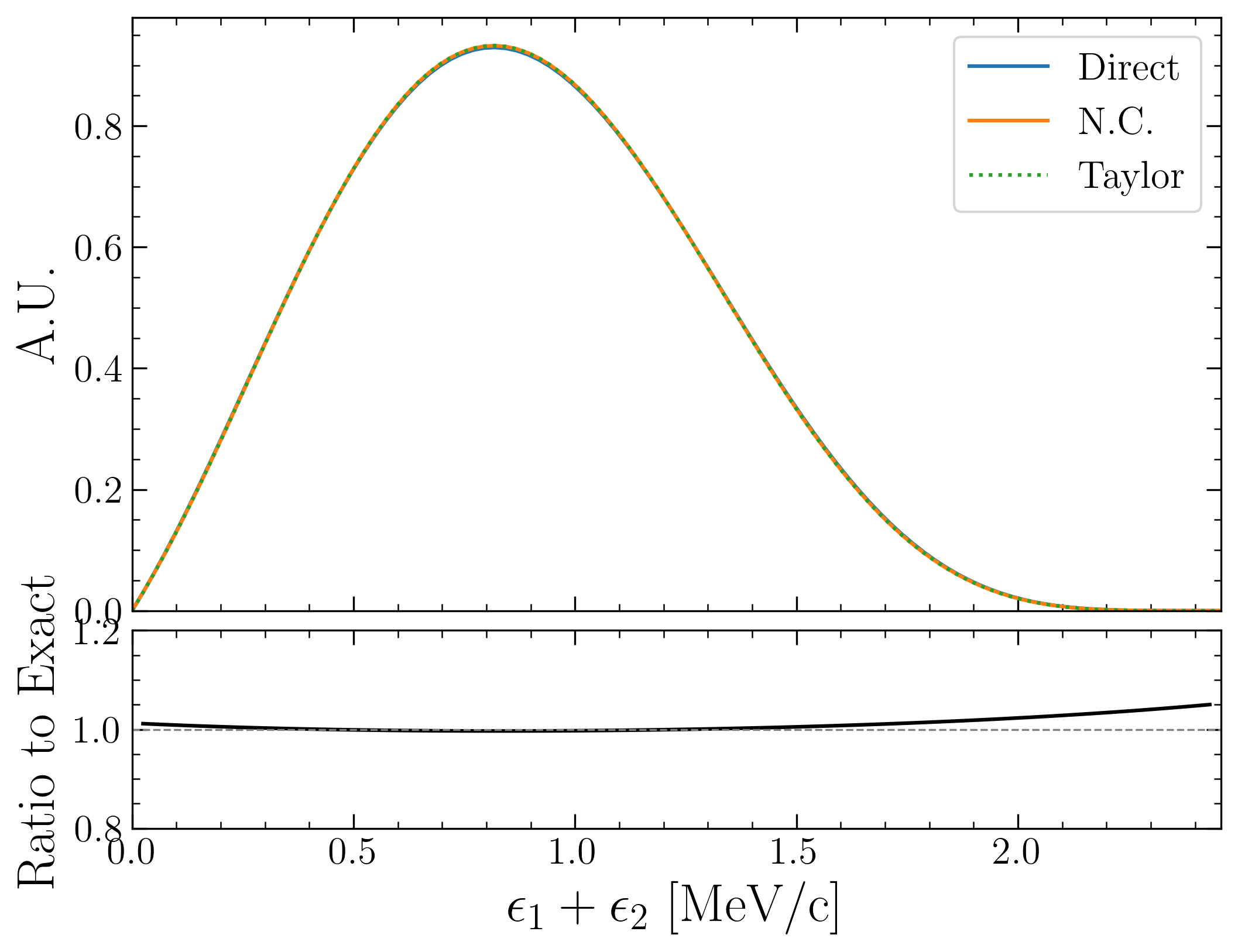}
    \caption{Same as in figure~\ref{fig:82Se_full_comp} but for $^{136}$Xe.}
    \label{fig:136Xe_full_comp}
\end{figure*}

\begin{figure*}[h!]
    \centering
    \includegraphics[width=0.45\linewidth]{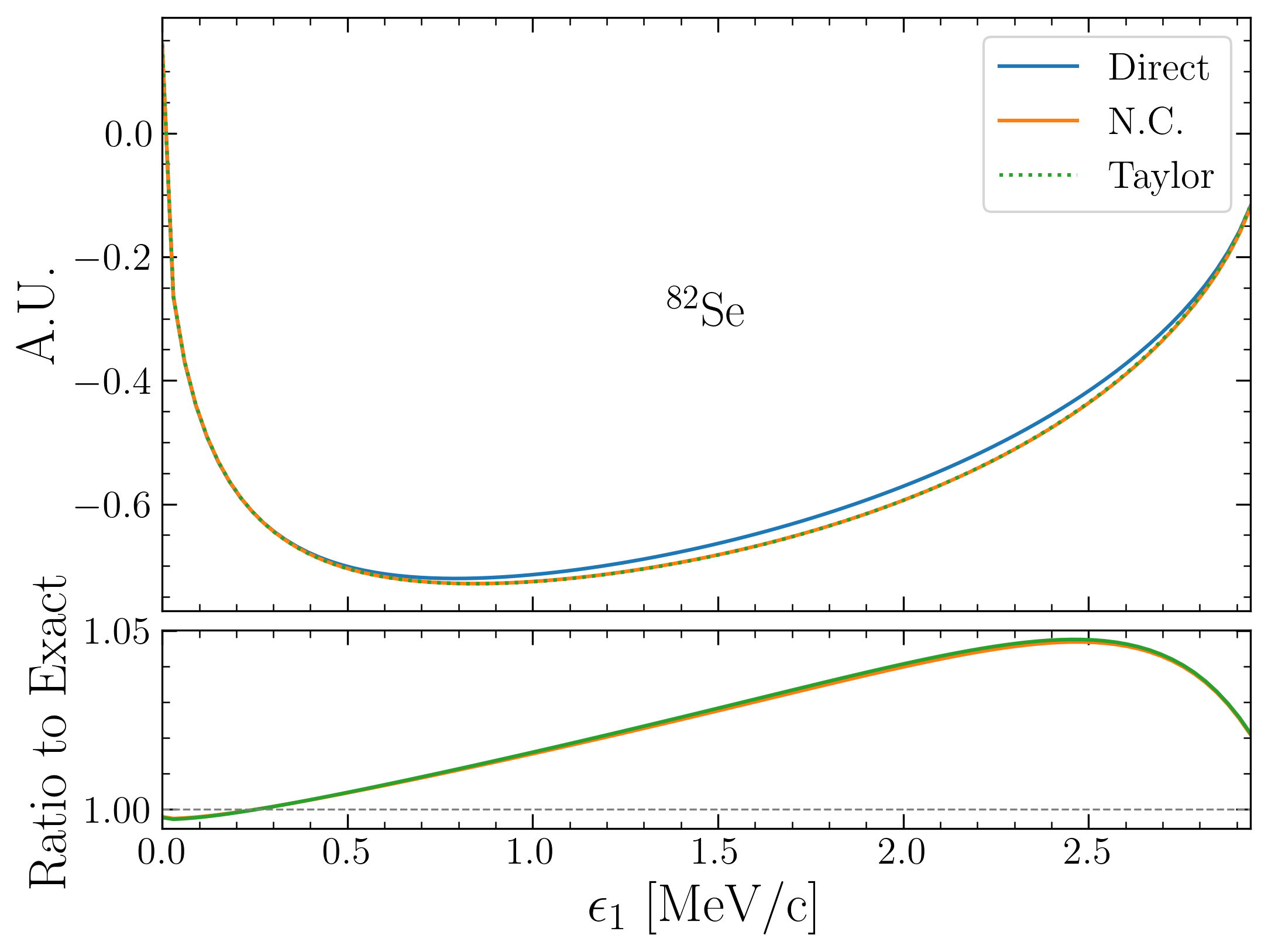}~
    \includegraphics[width=0.45\linewidth]{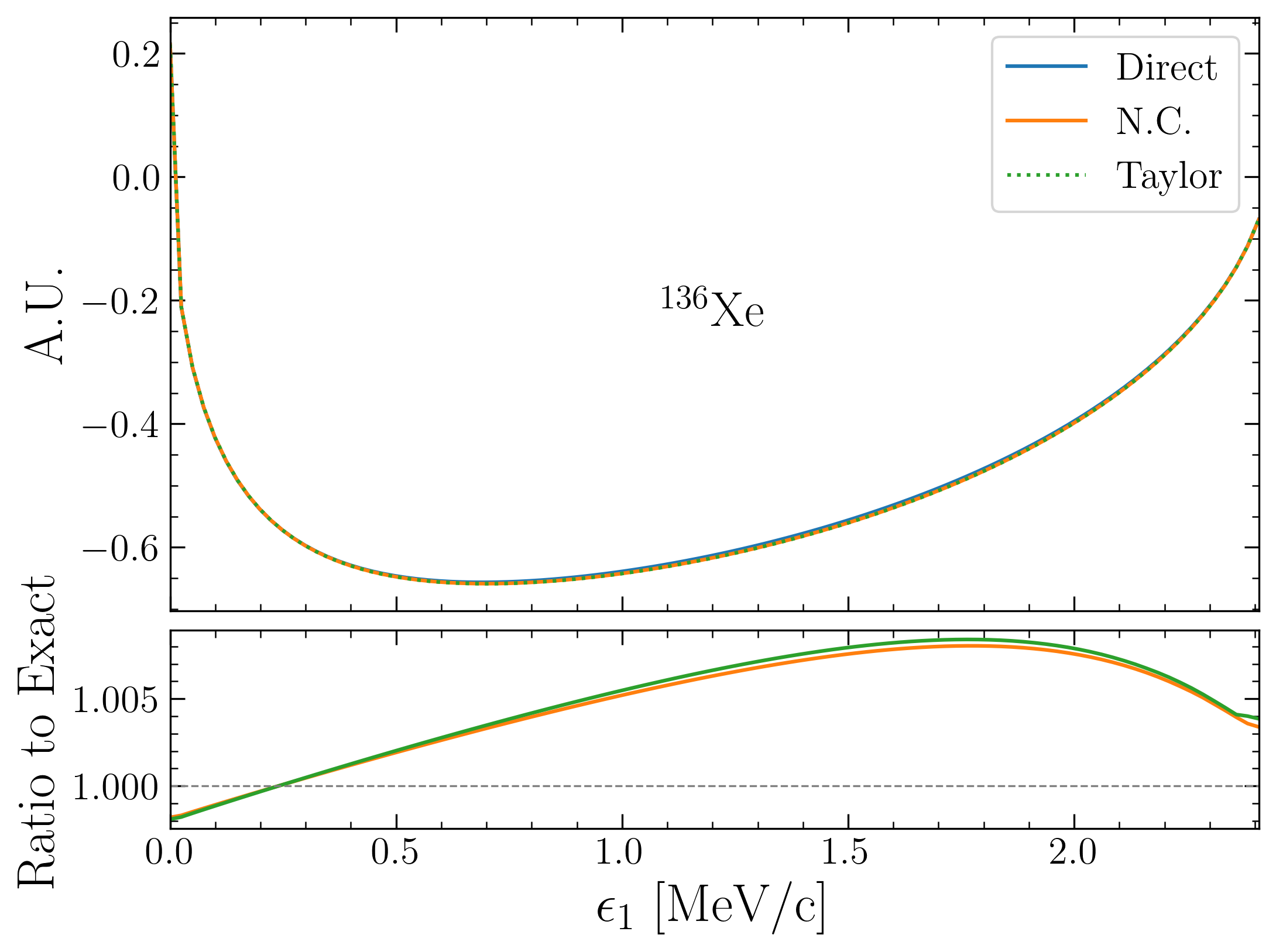}
    \caption{Angular correlations for $^{82}$Se (left) and $^{136}$Xe (right). The same conventions as in Fig.~\ref{fig:82Se_full_comp} are used.}
    \label{fig:ang_corr}
\end{figure*}

In this section, we present the calculations for the decay rates, electron spectra and angular correlations performed with our "Direct" method for the $^{82}$Se and $^{136}$Xe isotopes, and compare them with the values obtained with N.C. and Taylor methods. 
Our calculations are presented without employing any quenching of the NMEs, Gamow-Teller strength, or $g_A$ values. Additionally, we do not fit any parameters to the experimental results. This choice is deliberate, allowing us to focus on the differences between the approaches, independently of the use of different parameters in calculation adjusted to fit the experimental half-lives.
We calculated the $1^+$ spectrum of $^{82}$Br and the GT strengths required to compute the decay rate following the recipe of Ref.~\cite{Horoi-physics-2022} with the \verb|Antoine| nuclear shell model program~\cite{Antoine-SH} using the \verb|jun45|
~\cite{JUN45} effective Hamiltonian in the jj44 model space comprising of a $^{56}$Ni core and the $1p_{3/2}$, $1p_{1/2}$, $0f_{5/2}$, and~$0g_{9/2}$ valence orbitals. 
In this model space, one can perform calculations for nuclei between N,Z=28 and N,Z=50. However, this model space misses the $0f_{7/2}$ and the $0g_{7/2}$ orbitals, which are the spin-orbit partners of the $0f_{5/2}$ and~$0g_{9/2}$ orbitals, respectively, resulting in discrepancies between the experimental and the theoretical values of the nuclear matrix elements.
We have also assumed that the first excited $1^{+}$ state of $^{82}$Br is 75~keV above its ground state, as explained in~\cite{TULI2019260}. The energy of this level is well defined, however its spin and parity are assigned from theory. Nevertheless, the $1^+$ assignment for this level is common in DBD calculation and we employ it as well. For $^{136}$Xe, we also used \verb|Antoine|, but with the \verb|svd|~\cite{Chong2012} effective Hamiltonian in the $jj55$ model space consisting of a $^{100}$Sn core and a valence space of the $0g_{7/2}$, $1d_{5/2}$, $1d_{3/2}$, $2s_{1/2}$, and $0h_{11/2}$ orbitals, The spin-orbit partner orbitals of $0g_{7/2}$ and $0h_{11/2}$ are missing, leading in deviations from calculations in larger and computationally more challenging model spaces, such as jj77~\cite{Horoi-Brown_jj77_2013}.
The summation over the $1^+$ states encompassed 30 energy levels for $^{82}$Se and for $^{136}$Xe, which has been found to be sufficient for attaining convergence of the NME~\cite{horoi2026-arxiv}. For the N.C. calculations the values of $\langle E_{N}\rangle$ used in the PSFs are obtained from the usual approximation of the position of the giant Gamow-Teller resonance~\cite{Haxton-PPNP1984,Kotila_PRC_2012,Stoica_FP_2019}
\begin{equation*}
    \frac{Q_{\beta\beta}}{2}+m_e +\langle E_N\rangle - E_{I} = 1.12\times A^{1/2}.
\end{equation*}
The $Q_{\beta\beta}$-values are taken from~\cite{Ghinescu_EPJC_2026}, where they were computed based on atomic mass differences between the initial and final neutral atoms. The integrals above are computed with the \verb|SPADES| package~\cite{SPADES} as described in~\cite{Ghinescu_EPJC_2026}, except for the radiative corrections and exchange effects, which we will incorporate in a future paper.

The results for the decay rates are shown in Table~\ref{tab:half_lives_comp} in the form of half-lives. For both investigated nuclei, we obtain values that are up to 16\% smaller than the ones obtained with the approximate methods. An unambiguous conclusion on these differences cannot be drawn yet, without expanding the analysis to more isotopes and alternative methods of NME calculation. They may come from the interdependence between nuclear structure and lepton kinematics, explicitly taken into account by our method, but they may also be influenced by the nuclear model used. Consequently, evaluating how these discrepancies evolve when using different nuclear models within our formalism would be an interesting investigation. We also note here that the values presented are not meant to be compared with experimental ones. The purpose of Table~\ref{tab:half_lives_comp} is to compare computation methods using the same nuclear structure input.
In this context, most of the differences between experimental values and the calculated half-lives can be attributed to the missing spin-orbit partner orbitals and from our choice of not using any quenching factors on the values of the nuclear matrix elements.

Next, we calculated the single and summed energy electron spectra as well as the angular correlation spectra and show the difference between the Direct method and the approximate ones using the full $1^+$ spectrum, for the isotopes $^{82}$Se and $^{136}$Xe in Figures~\ref{fig:82Se_full_comp},~\ref{fig:136Xe_full_comp} and~\ref{fig:ang_corr} respectively. We note that the spectra computed in the Taylor approximation contain only the first order term. This is consistent with past studies~\cite{Simkovic_PRC_2018,KamLANDZen_PRL_2019}.

As seen from all three figures, the single electron spectra are most sensitive to highlighting the differences between the Direct and approximate methods of calculation. For $^{82}$Se, we observe differences of up to 20\% between methods, which are within the experimentally accessible range of the single electron spectrum. The approximate summed electron energy spectra are within 10\% of the ones computed with the Direct method, for this nuclide, while the differences in the angular correlation spectra are of at most 5\%. The differences between the approximate methods and the Direct one are significantly smaller in the case of $^{136}$Xe. This is expected since $E_{N}-\frac{1}{2}(E_I+E_F)$ sits at $\simeq$ 680~keV above $Q_{\beta\beta}$, making both the N.C. and Taylor approximations quite accurate. 
Finally, we remark that our method is well suited for exploring the Single-State Dominance (SSD) and Higher-State Dominance (HSD) hypotheses regarding the contribution of the virtual $1^+$ states to the total decay amplitude. To facilitate experimental investigations, we can provide detailed datasets of our calculations upon request.

\section{Conclusions}
In conclusion, we proposed a novel method to calculate the decay rate and electron spectra which offers a more realistic description of the $2\nu\beta\beta$ decay process, by preserving the complete interdependence between nuclear structure and lepton kinematics.    
Our Direct calculations reveal increases of up to $16\%$ in the decay rates for the investigated isotopes, $^{82}\text{Se}$ and $^{136}\text{Xe}$, compared to traditional approximate methods, resulting in shorter predicted half-lives. 
We also calculate the electron spectra within the same unified framework. While the summed electron energy spectra remain largely insensitive to the calculation method, {\color{blue} we point out that the} single-electron spectra exhibit sensitivity within the experimentally accessible range, particularly for $^{82}\text{Se}$. The effects of the nuclear structure on the angular correlation spectra lie somewhere in between the aforementioned cases. This sensitivity could prove a valuable experimental tool for validating nuclear structure calculations, SSD or HSD hypotheses, and probing the subtle interdependence between nuclear structure and lepton kinematics in DBD. 

\backmatter
\bmhead{Acknowledgements}
We acknowledge support from project PNRR-I8/C9-CF264, Contract No. 760100/23.05.2023 of the Romanian National Authority for Research.

\bibliography{bibliography}
\end{document}